\documentclass[a4paper,11pt]{article}
\usepackage[utf8]{inputenc}
\usepackage[DIV=12]{typearea}
\usepackage{graphicx}
\usepackage{amsmath,amsthm,amssymb}
\usepackage[mathlines,pagewise]{lineno}
\usepackage{listings}
\usepackage[draft]{varioref}
\usepackage{caption}
\usepackage{url}
\usepackage{todonotes}

\renewcommand{\epsilon}{\varepsilon}

\newcommand{\MF}{\mathcal{MF}}
\newcommand{\MAB}{\mathit{MAB}}

\newcommand{\Bal}{\mathit{Bal}}

\newtheorem{theorem}{Theorem}
\newtheorem{proposition}[theorem]{Proposition}
\newtheorem{lemma}[theorem]{Lemma}
\newtheorem{corollary}[theorem]{Corollary}

\theoremstyle{definition}
\newtheorem{definition}[theorem]{Definition}
\newtheorem{example}[theorem]{Example}
\newtheorem{problem}[theorem]{Problem}
\newtheorem{remark}[theorem]{Remark}

\title{Some Results on Digital Segments and Balanced Words}
\author{Alessandro De Luca \and Gabriele Fici}
\date{}

\begin{document}

\maketitle

\begin{abstract}
We exhibit combinatorial results on Christoffel words and binary balanced words that are motivated by their geometric interpretation as approximations of digital segments. 
We give a closed formula for counting the exact number of balanced words with $a$ zeroes and $b$ ones. We also study minimal non-balanced words.\\[2ex]
\end{abstract}

\section{Introduction}

The study of digital approximations of lines in the plane by discrete paths in the grid $\mathbb{Z}^2$ is a classical topic at the frontier between digital geometry, arithmetic, and combinatorics on words. For example, it is well known that digital (infinite) rays with irrational slope are encoded by infinite \emph{Sturmian words}. A word is a sequence of letters drawn from a finite alphabet. A (right) infinite word is Sturmian if it has $n+1$ distinct factors (contiguous sub-blocks) of length $n$ for every $n\geq 0$. Since the famous theorem of Morse and Hedlund~\cite{MoHe38} states that a right infinite word is aperiodic if and only if it has \emph{at least} $n+1$ distinct factors  of length $n$ for every $n\geq 0$, Sturmian words are aperiodic binary words of minimal factor complexity (see~\cite[Chap.~2]{LothaireAlg} for more details). 

A geometric interpretation of  Sturmian words can be given in terms of \emph{cutting sequences}. Consider rays of the form $y = \alpha x + \rho$ with irrational slope $\alpha>0$ and real intercept $\rho\geq 0$. 
The cutting sequences of such lines are defined by labeling the intersections of $y = \alpha x +\rho$ with the (first quadrant of the) integer grid, using $0$ if the grid line crossed is vertical and $1$ if it is horizontal (and either $01$ or $10$ if the ray passes through a point of $\mathbb Z^2$). The sequence of labels, read from $(0,\rho)$ out, is called the cutting sequence of $y = \alpha x+\rho$ and is a Sturmian word.
%


Among many equivalent definitions, Sturmian words are also characterized by the \emph{balance property}. A (finite or infinite) binary word is balanced if in every two factors of the same length, the difference between occurrences of each letter is bounded by $1$. Sturmian words are precisely aperiodic balanced binary words~\cite{MoHe40}. 

The case of finite segments is also source of a large number of interesting correspondences between digital geometry, arithmetic and combinatorics. For a general overview, see~\cite{survey}. In particular, finite factors of Sturmian words are precisely the binary balanced words. Among these, a special role is played by \emph{Christoffel words}. The literature about Christoffel words is very rich (see for example \cite{DBLP:journals/tcs/Reutenauer15,LothaireAlg,JTNB_1993}).  A lower (resp., upper) Christoffel word is a binary word  that encodes the digital approximation from below (resp., from above) of a Euclidean segment, where $0$ (resp., $1$) represents a horizontal (resp., vertical) unit segment~(see Fig.~\ref{fig:chr}). 

For every pair of positive integers $(a,b)$ there is one lower (resp.~upper) Christoffel word having $a$ occurrences of $0$ and $b$ occurrences of $1$. These two words define the contour (lower and upper, respectively) of a region enclosing the Euclidean segment from $(0,0)$ to $(a,b)$ (see~Fig.~\ref{fig:stripe}).

We give a purely combinatorial proof of the fact  that \emph{every} balanced word with $a$ occurrences of $0$ and $b$ occurrences of $1$ encodes a path that is contained in this region. 

So, given positive $a$ and $b$, the set of balanced words with $a$ occurrences of $0$ and $b$ occurrences of $1$ is in general a proper subset of the set of binary words encoding a path that lies at distance bounded by $\sqrt{2}$ from the Euclidean segment joining the origin to the point $(a,b)$.
A natural question that therefore arises is that of counting, for every pair of positive integers $(a,b)$, the exact number of balanced words with $a$ occurrences of $0$ and $b$ occurrences of $1$. While a recursive formula for this problem was known~\cite{DBLP:journals/dmtcs/BedarideDJR10},
in this paper we present an exact closed formula.

We  also study the set of minimal non-balanced words and the set of minimal almost balanced words, providing new combinatorial characterizations for these classes of words already studied in \cite{DBLP:journals/tcs/Provencal11}.

\section{Preliminaries}\label{sec:prel}

 A \emph{word} over a finite alphabet $\Sigma$ is a concatenation of letters from $\Sigma$. The \emph{length} of a word $w$ is denoted by $|w|$. The empty word $\varepsilon$ has length $0$. The set of all words 
 over  $\Sigma$ is denoted  $\Sigma^*$
 and is a free monoid
 with respect to concatenation. 
 For a letter $x\in\Sigma$, $|w|_x$ denotes the number of occurrences of $x$ in $w$. If $\Sigma=\{x_1,\ldots,x_\sigma\}$ is ordered, the vector $(|w|_{x_1},\ldots,|w|_{x_\sigma})$ is the \emph{Parikh vector} of $w$.

Let $w=uv$, with $u,v\in \Sigma^*$. We say that $u$ is a \emph{prefix} of $w$ and that $v$ is a \emph{suffix} of $w$. A \emph{factor} of $w$ is a prefix of a suffix (or, equivalently, a suffix of a prefix) of $w$. 
If $w=w_1w_2\cdots w_n$, with $w_k\in\Sigma$ for all $k$, we let $w[i..j]$ denote the nonempty factor $w_iw_{i+1}\cdots w_j$ of $w$, whenever $1\leq i\leq j\leq n$.
A factor $u$ of a word $w\neq u$ is a \emph{border} of $w$ if $u$ is both a prefix and a suffix of $w$; in this case, $w$ has \emph{period} $|w|-|u|$. A word $w$ is \emph{unbordered} if its longest border is $\varepsilon$; i.e., if its smallest period is $|w|$. The prefix $\rho_w$ whose length equals the smallest period of a nonempty $w$ is called its \emph{fractional root}.
For a word $w$, the \emph{$n$-th power} of $w$ is the word $w^n$ obtained by concatenating $n$ copies of $w$; furthermore, we let $w^\omega$ denote the infinite periodic word $www\cdots$.

Two words $w$ and $w'$ are \emph{conjugate} if $w=uv$ and $w'=vu$ for some words $u$ and $v$. The conjugacy class of a word $w$ contains $|w|$ elements if and only if $w$ is \emph{primitive}, i.e., $w=v^n$ for some $v$ implies $n=1$.

A nonempty word $w=w_1w_2\cdots w_n$, $w_k\in \Sigma$ for all $k$, is a \emph{palindrome} if  it coincides with its \emph{reversal} $w^R=w_nw_{n-1}\cdots w_1$. The empty word is also assumed to be a palindrome. 

The following proposition, whose proof is straightforward, is well known (cf.~\cite{DELUCA1982207,DBLP:journals/ijfcs/BrlekHNR04}). 

\begin{proposition}\label{prop:2pal}
 A word is a conjugate of its reversal if and only if it is a concatenation of two palindromes. Moreover, these two palindromes are uniquely determined if and only if the word is primitive.
\end{proposition}
 


\subsection{Christoffel Words}\label{sec:chris}

We now define Christoffel words and their properties. We point the reader to the classical references~\cite{Book08,ReutenauerMarkoff,LothaireAlg,Be07} for more details on Christoffel words.

In what follows, we fix the alphabet $\{0,1\}$ and represent a word over $\{0,1\}$ as a path in $\mathbb{N}\times \mathbb{N}$ starting at $(0,0)$, where $0$ (resp., $1$) encodes a horizontal (resp., vertical) unit segment.
For every pair of nonnegative integers $(a,b)$ (not both $0$), every word $w$ that encodes a path from $(0,0)$ to $(a,b)$ must have exactly $a$ zeroes and $b$ ones, i.e., it has Parikh vector $(a,b)=(|w|_0,|w|_1)$. We also say that $b=|w|_1$ is the \emph{height} of $w$. 
We define the \emph{slope} of a word $w$ with Parikh vector $(a,b)$ as the rational number $b/a$ if $a\neq 0$, or $\infty$ otherwise.
We remark here that some authors~(see, e.g.,~\cite{LothaireAlg}) call slope the ratio $b/(a+b)$ between the height and the length of the word; in this paper we will refer to this quantity as the \emph{mechanical slope}.

\begin{definition} 
Given a pair of nonnegative integers $(a,b)$ (not both $0$), the \emph{lower (resp.,~upper) Christoffel word} $w_{a,b}$ (resp.,~$W_{a,b}$) is the word encoding the path from $(0,0)$ to $(a,b)$ that is closest from below (resp.,~from above) to the Euclidean segment, without crossing it. 
 \end{definition}

\begin{figure}[ht]
     \centering
     \includegraphics[width=70mm]{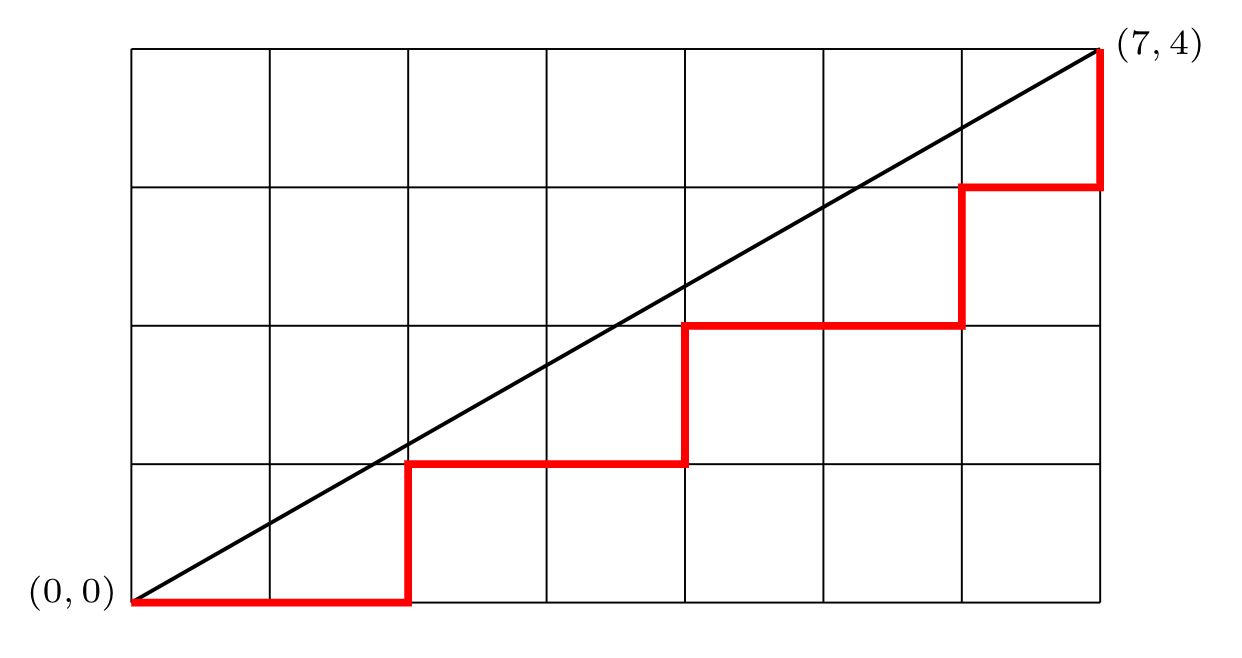}
     \caption{The lower Christoffel word $w_{7,4}=00100100101$. The upper Christoffel word $W_{7,4}=10100100100$ is the reversal of $w_{7,4}$.}
     \label{fig:chr}
 \end{figure}

 In other words, $w_{a,b}$ (resp.~$W_{a,b}$) is the digital approximation from below (resp., from above) of the Euclidean segment joining $(0,0)$ to $(a,b)$.
 For example, the lower Christoffel word $w_{7,4}$ is the word $00100100101$ (see~Fig.~\ref{fig:chr}).
 
 By construction, we have that the upper Christoffel word $W_{a,b}$ is the reversal of the lower Christoffel word $w_{a,b}$ (and the two words are conjugates, see below).

 Notice that by definition we allow $a$ or $b$ (but not both) to be $0$, that is, we have $w_{n,0}=W_{n,0}=0^n$ and $w_{0,n}=W_{0,n}=1^n$; the words of length $1$ are both lower and upper Christoffel words.

If $a$ and $b$ are coprime, the Christoffel words $w_{a,b}$ and $W_{a,b}$ do not intersect the Euclidean segment joining $(0,0)$ to $(a,b)$ (other than at the end points) and are primitive words. If instead $a=n\alpha$ and $b=n\beta$ for some $n>1$, then $w_{a,b}=(w_{\alpha,\beta})^n$ (resp.,~$W_{a,b}=(W_{\alpha,\beta})^n$).  Hence, $w_{a,b}$ (resp.,~$W_{a,b}$) is primitive if and only if $a$ and $b$ are coprime.

Therefore, for every $n> 1$, there are $2\phi(n)$ primitive Christoffel words of length $n$ (in particular, there are $\phi(n)$ primitive lower Christoffel words and $\phi(n)$ primitive upper Christoffel words), where $\phi$ is the Euler totient function.

\begin{figure}
    \centering
    \includegraphics[width=280px]{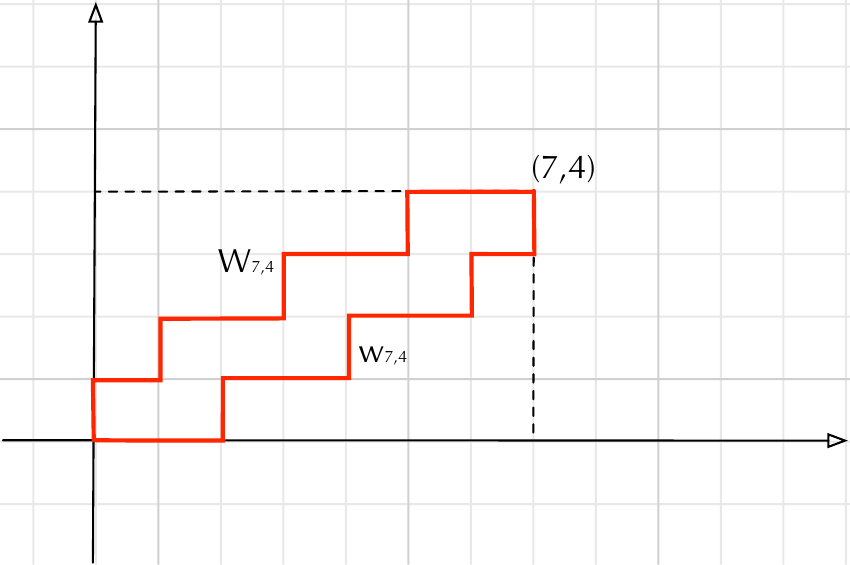}
    \caption{The region $R(7,4)$ of the discrete plane whose contour is delimited by the lower Christoffel word $w_{7,4}$ and the upper Christoffel word $W_{7,4}$ is precisely the region of points in the grid that have Euclidean distance smaller than $\sqrt{2}$ from the Euclidean segment joining $(0,0)$ to $(7,4)$.}
    \label{fig:stripe}
\end{figure}

 \begin{remark}\label{rem:new}
From the definition, it follows that 
the region $R(a,b)$ of the discrete plane whose contour is delimited by the lower and the upper Christoffel word with Parikh vector $(a,b)$ is precisely the region of points in the grid that have Euclidean distance smaller than $\sqrt{2}$ from the Euclidean segment joining $(0,0)$ to $(a,b)$, see Fig.~\ref{fig:stripe}. Notice that by definition, this region does not contain internal points with integer coordinates.
\end{remark}

Consider the sequence $(k\cdot b \bmod (a+b))$, for $k=1,\ldots,a+b$. Each subsequent number in the sequence is obtained either by adding $b$ (supposing to start from $0$) or by subtracting $a$. Writing $0$ whenever we add $b$ and $1$ when we subtract $a$ we get the lower Christoffel word $w_{a,b}$ (as is well known and easy to see from the definition). If we divide each term in the sequence by $a$, we get the sequence of vertical distances between the endpoints of paths encoded by prefixes of  $w_{a,b}$ and the Euclidean segment joining $(0,0)$ to $(a,b)$; if instead we divide by $\sqrt{a^2+b^2}$  
(the length of the Euclidean segment) we get the sequence of Euclidean distances. In particular, the point with integer coordinates that is closest to the Euclidean segment (without lying on the segment itself) is at distance $1/\sqrt{a^2+b^2}$, whereas the point on the path representing $w_{a,b}$ that is \emph{farthest} from the segment is always at distance $(a+b-1)/\sqrt{a^2+b^2}$.
These two points play a crucial role in the combinatorics of Christoffel words (see Sec.~\ref{sec:fact}).

For example, for the path $w_{7,4}$ depicted in Figure~\ref{fig:chr}, the sequence $(4k\bmod 11)$ is
\[4,8,1,5,9,2,6,10,3,7,0.\]  Taking for instance $k=5$, we get that the point $(4,1)$, which is the endpoint of the path corresponding to the prefix $00100$ of $w_{7,4}$, has vertical distance $9/7\approx 1.286$ and Euclidean distance $9/\sqrt{65}\approx 1.116$ from the Euclidean segment joining $(0,0)$ and $(7,4)$.

%

We now give a fundamental definition for the combinatorics of Christoffel words.

\begin{definition}
 A \emph{central word} is a word that has coprime periods $p$ and $q$ and length $p+q-2$.
\end{definition}

It is well known that central words are binary palindromes~\cite{DBLP:journals/tcs/LucaM94,LothaireAlg}. The following theorem gives a structural characterization (see~\cite{DBLP:journals/tcs/Luca97a}).

\begin{theorem}[Combinatorial Structure of Central Words]\label{thm:CSCW}
 A word $w$ is central if and only if it is a power of a letter or there exist palindromes $P$ and $Q$ such that $w=P01Q=Q10P$.
 Moreover,
\begin{itemize}
\item $P$ and $Q$ are central words;
\item $|P|+2$ and $|Q|+2$ are coprime and $w$ has periods $|P|+2$ and $|Q|+2$;
\item if $|P|<|Q|$, $Q$ is the longest palindromic (proper) suffix of $w$.
\end{itemize}
\end{theorem}

\begin{example}
The word $w=010010$ is a central word, since it has periods $3$ and $5$ and length $6=3+5-2$. We have $010010=010 \cdot 01 \cdot 0=0 \cdot 10 \cdot 010$, so that $P=010$ and $Q=0$.
\end{example}
 
\begin{proposition}[\cite{Pirillo2001}]\label{prop:paldec}
A word $C$ is central if and only if the words $0C1$ and $1C0$ are conjugates.
\end{proposition}
 

\begin{proposition}[\cite{DBLP:journals/tcs/BerstelL97}]
 A word is a primitive lower (resp.,~upper) Christoffel  word if and only if it has length $1$ or it is of the form $0C1$ (resp.,~$1C0$) where $C$ is a central word.
\end{proposition}

\begin{figure}[ht]
     \centering
    \includegraphics[width=70mm]{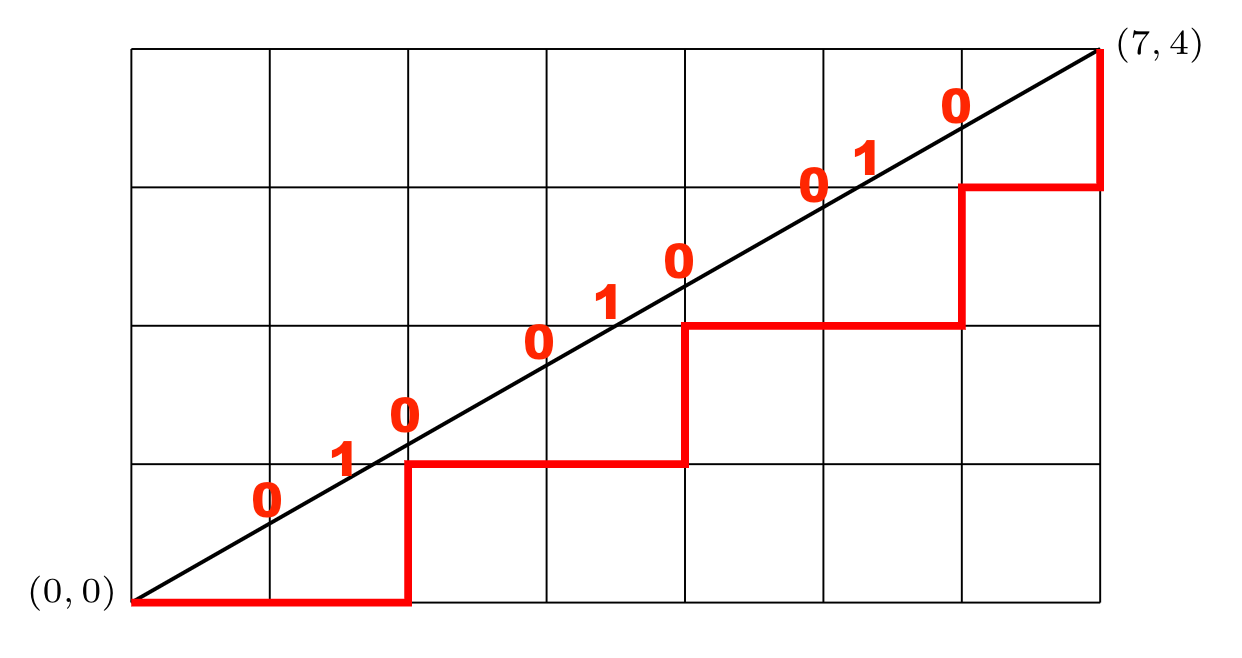}
     \caption{The central word $C=010010010$ is the central factor of the primitive lower Christoffel  word $w_{7,4}=0C1$. It encodes the intersections of the Euclidean segment joining $(0,0)$ and $(7,4)$ with the grid ($0$ for a vertical intersection and $1$ for a horizontal intersection).}
     \label{fig:central}
 \end{figure}

\begin{example}
 Let $a=7$ and $b=4$. We have  $w_{7,4}=00100100101=0\cdot 010010010 \cdot 1$, where $~C=010010010$ is a central word since it has periods $3$ and $8$ and length $9=3+8-2$; see Fig.~\ref{fig:central}.
\end{example}

So, central words are the ``central'' factors of primitive Christoffel  words of length~$\geq 2$. 
A geometric interpretation of the central word $C$ is the following: it encodes the intersections of the Euclidean segment joining $(0,0)$ to $(a,b)$ ($0$ for a vertical intersection and $1$ for a horizontal intersection). In other words, the word $C$ is the cutting sequence of the Euclidean segment joining $(0,0)$ to $(a,b)$; see again Fig.~\ref{fig:central}.

\begin{proposition}[\cite{DBLP:journals/ejc/BertheLR08}]\label{prop:multinv}
Let $w_{a,b}=0C1$ be a primitive lower Christoffel word. The central word $C$ has periods $a'$ and $b'$, the multiplicative inverses of $a$ and $b$ modulo $a+b$, respectively.
\end{proposition}



\begin{definition}
 A word over $\{0,1\}$ is \emph{balanced} if the number of $0$s (or, equivalently, $1$s) in every two factors of the same length differs by at most $1$.
\end{definition}

We let $\Bal$ denote the set of balanced words over the alphabet $\{0,1\}$. 

\begin{remark}\label{rem:fact}
 Balance is a \emph{factorial property}, i.e., every factor of a balanced word is itself balanced.
\end{remark}

Binary balanced words are precisely the finite factors of Sturmian words. 

From the definition of balanced word, it immediately follows  that if $w$ is a binary balanced word, then for every $1\leq k\leq |w|$, the factors of $w$ of length $k$ can have at most two different Parikh vectors: $(i,k-i)$, $(i-1,k-i+1)$, for some $0\leq i\leq k$. According to~\cite{Rigo}, we call the factors of the first kind \emph{light factors} and those of the second kind \emph{heavy factors}. By convention, if all factors of length $k$ have the same Parikh vector, we consider them light. 

For example, the word $1011010$ is balanced; its factors of length $3$ are: $010$, $011$, $101$, $110$. The first one is the only light factor of length $3$, the other ones are heavy.

 Notice that in a primitive lower Christoffel word $w_{a,b}=0C1$, all prefixes are light factors and all proper suffixes are heavy. Indeed, if $0C1$ is a primitive lower Christoffel word
 then since $C$ is a palindrome, for every prefix $0v$ of $w_{a,b}$ we have that the suffix of the same length is $v^R1$, and clearly $v$ and $v^R$ have the same Parikh vector. The situation is of course symmetric for primitive upper Christoffel words.

\begin{theorem}[\cite{DBLP:journals/tcs/BerstelL97}]\label{thm:Lyndon}
Let $w$ be a  word over $\{0,1\}$. Then $w$ is a primitive (lower or upper) Christoffel word if and only if it is balanced and unbordered.
\end{theorem}

In particular, the set of primitive lower Christoffel  words is precisely the set of balanced \emph{Lyndon words} over the alphabet $\{0,1\}$  (for the order $0<1$). 
 A Lyndon word is a primitive word that is lexicographically smaller than all its conjugates (or, equivalently, lexicographically smaller than all its proper suffixes).

\begin{proposition}[\cite{JTNB_1993}]\label{prop:BL}
 For every coprime $a,b$, the primitive lower Christoffel word $w_{a,b}$ is the greatest (for the lexicographic order) Lyndon word among all Lyndon words having Parikh vector $(a,b)$.
\end{proposition}

\subsection{Factorizations of Christoffel Words}\label{sec:fact}

A basic result in the theory of Lyndon words is that every Lyndon word $w$ of length $|w|\geq 2$ has a \emph{standard factorization} $w=uv$, where $v$ is the lexicographically least proper suffix of $w$ (or, equivalently, the longest proper suffix of $w$ that is a Lyndon word), see~\cite{Lot01}. Since primitive lower Christoffel words are Lyndon words, every primitive lower Christoffel word of length $|w|\geq 2$ has a standard factorization. 

On the other hand, by Proposition~\ref{prop:paldec}, a primitive lower Christoffel word is a conjugate of its reversal (the corresponding upper Christoffel word); hence by Proposition~\ref{prop:2pal}, every primitive lower Christoffel word of length $|w|\geq 2$ has a unique \emph{palindromic factorization}.

Let $w_{a,b}=0C1$ be a primitive lower Christoffel word, so that $a$ and $b$ are coprime integers. If the central word  $C$ is not a power of a single letter, then by Theorem~\ref{thm:CSCW} there exist central words $P$ and $Q$ such that $C=P01Q=Q10P$ so that $w_{a,b}=0C1=0P0\cdot 1Q1=0Q1\cdot 0P1$. 

Hence, we have the following factorizations:

\begin{enumerate}
 \item $0C1=0P0\cdot 1Q1$ (palindromic factorization);
 \item $0C1=0Q1\cdot 0P1$ (standard factorization).
\end{enumerate}


If instead $C=0^n$ (the case $C=1^n$ is analogous)  we have:
\begin{enumerate}
 \item $0C1=0^{n+1}\cdot 1$ (palindromic factorization);
 \item $0C1=0\cdot 0^n1$ (standard factorization).
\end{enumerate}

Actually, for every lower Christoffel word $w_{a,b}$ (not necessarily primitive), the following holds: Let $P$, $P'$ be two palindromes such that $w_{a,b}=PP'$. Then $P'P=w^R_{a,b}=W_{a,b}$. 

\begin{remark}\label{rem:periods}
 Notice that the lengths of the two factors in both the palindromic and the standard factorization of the primitive lower Christoffel word $w_{a,b}=0C1$ are precisely the two coprime periods of the central word $C$ whose sum is $a+b$, i.e., by Proposition~\ref{prop:multinv}, the multiplicative inverses of $a$ and $b$ modulo $a+b$.
\end{remark}

\begin{figure}
\begin{center}
\begin{minipage}{7.2cm}
\includegraphics[width=70mm]{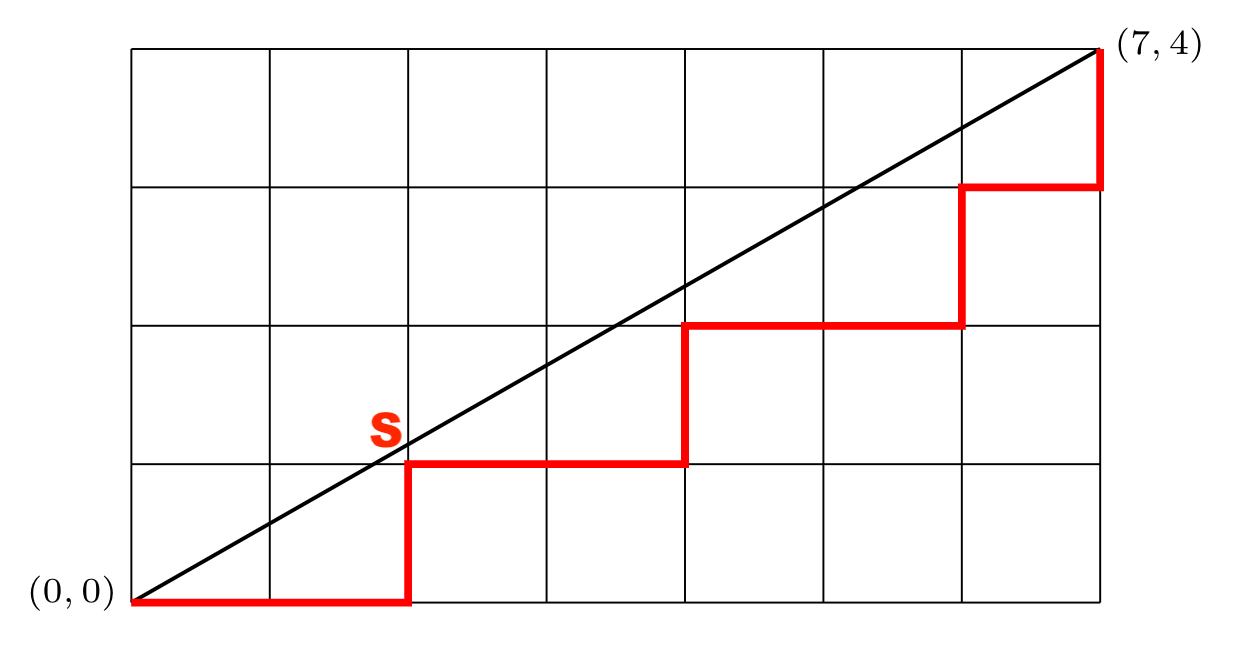}
\end{minipage}
\begin{minipage}{7.2cm}
  \includegraphics[width=70mm]{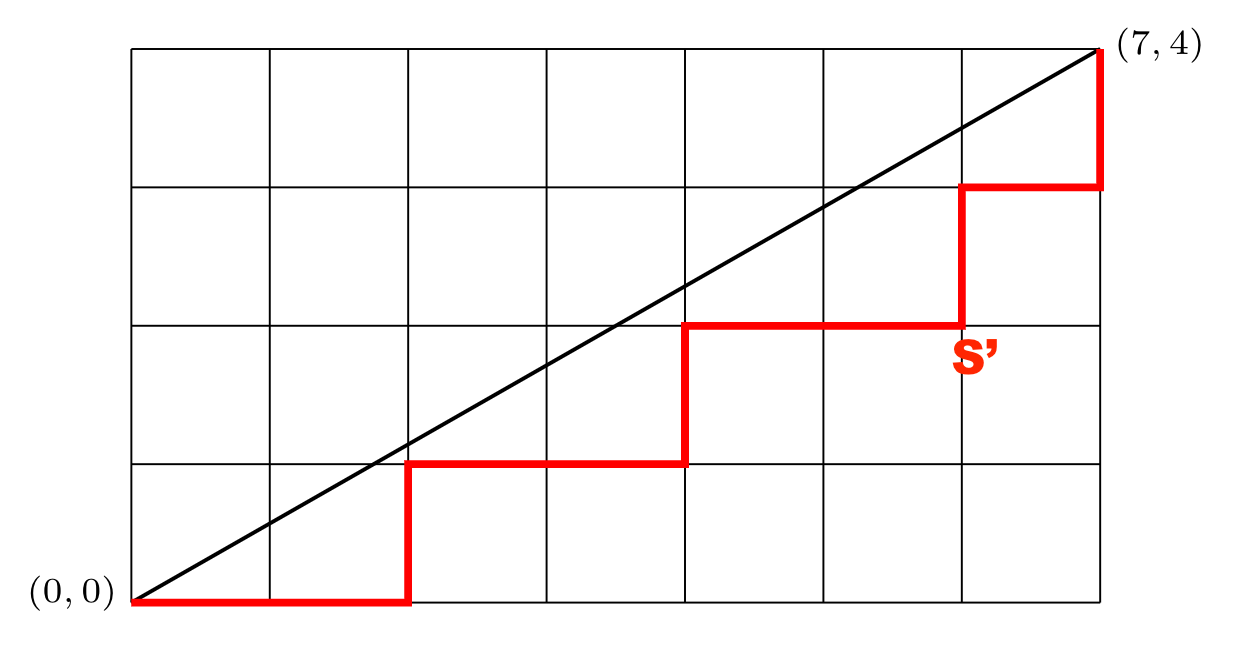}
\end{minipage}
\end{center}
    \caption{The standard factorization $0Q1\cdot 0P1 = 001 \cdot 00100101$ (left) and the palindromic factorization $0P0\cdot 1Q1 = 00100100\cdot 101$ (right) of the lower Christoffel word $w_{7,4}$. The point $S$ determined by the standard factorization is the closest to the Euclidean segment, while the point $S'$ determined by the palindromic factorization is the farthest. \label{fig:dec}}
\end{figure}

From the geometric point of view, the standard factorization divides $w_{a,b}$ in two shorter Christoffel words, and it determines the point $S$ of the encoded path that is \emph{closest} to the Euclidean segment from $(0,0)$ to $(a,b)$; the palindromic factorization, instead, divides $w_{a,b}$ in two palindromes and determines the point $S'$ that is \emph{farthest} from the Euclidean segment (see~\cite{DBLP:journals/ita/BorelR06,LamaPhD}). An example is given in~Fig.~\ref{fig:dec}.  

\begin{remark}
In the lower Christoffel word $(0C1)^2=0C1\cdot 0C1$ we have precisely one occurrence as a factor of the upper primitive Christoffel word $1C0$. Using the palindromic factorization, we can write \[(0C1)^2=0P0\cdot 1Q10P0 \cdot 1Q1=0P0\cdot 1C0 \cdot 1Q1\] (the case when $C$ is a power of a letter is analogous); so, the occurrence of $1C0$ in $(0C1)^2$ begins precisely in the point $S'$ (see~\cite{DBLP:journals/ita/BorelR06}).
\end{remark}

\subsection{Digital Approximations of a Given Euclidean Segment}

Fix a point $(a,b)$ in the grid ($a$ and $b$ nonnegative integers, not both $0$) and consider all the binary words that encode a path from $(0,0)$ to $(a,b)$, i.e., all the words that have Parikh vector $(a,b)$.
We aim at giving a combinatorial interpretation of those paths that better approximate the Euclidean segment from $(0,0)$ to $(a,b)$ according to some definition of digital approximation. Among these, there are the paths encoded by the conjugates of the lower and of the upper Christoffel word, as we are showing below. 


Let $a,b>0$ and $n=a+b$. The \emph{Christoffel matrix} $\mathcal{A}_{a,b}$ is the $n\times n$ matrix  in which the first column is a block of $a$ $0$'s followed by a block of $b$ $1$'s, and every subsequent column is obtained by shifting up the block of $1$'s by $b$ positions, modulo $n$, see~\cite{DBLP:journals/tcs/JenkinsonZ04,DBLP:journals/ita/BorelR06}. An example is given in Table~\ref{tab:ChrisArr}.


\begin{table}[ht]
\begin{center}
$$
\mathcal{A}_{7,4}=
\left(
\begin{matrix}
0 & 0 & 1 & 0 & 0 & 1 & 0 & 0 & 1 & 0 & 1 \\ 
0 & 0 & 1 & 0 & 0 & 1 & 0 & 1 & 0 & 0 & 1 \\ 
0 & 0 & 1 & 0 & 1 & 0 & 0 & 1 & 0 & 0 & 1 \\ 
0 & 1 & 0 & 0 & 1 & 0 & 0 & 1 & 0 & 0 & 1 \\ 
0 & 1 & 0 & 0 & 1 & 0 & 0 & 1 & 0 & 1 & 0 \\ 
0 & 1 & 0 & 0 & 1 & 0 & 1 & 0 & 0 & 1 & 0 \\ 
0 & 1 & 0 & 1 & 0 & 0 & 1 & 0 & 0 & 1 & 0 \\ 
1 & 0 & 0 & 1 & 0 & 0 & 1 & 0 & 0 & 1 & 0 \\ 
1 & 0 & 0 & 1 & 0 & 0 & 1 & 0 & 1 & 0 & 0 \\ 
1 & 0 & 0 & 1 & 0 & 1 & 0 & 0 & 1 & 0 & 0 \\ 
1 & 0 & 1 & 0 & 0 & 1 & 0 & 0 & 1 & 0 & 0 
\end{matrix}
\right) $$
\end{center}
\caption{The Christoffel matrix $\mathcal{A}_{7,4}$.\label{tab:ChrisArr}}
\end{table}

The first row of the Christoffel matrix $\mathcal{A}_{a,b}$ is the lower Christoffel word $w_{a,b}$. 
Every subsequent row is obtained from the previous one by swapping a $01$ factor with $10$. The last row is the upper Christoffel word $W_{a,b}$.
Actually, the rows are precisely the conjugates of $w_{a,b}$ and appear in lexicographic order. The rows are all distinct if and only if $w_{a,b}$ is primitive, i.e., $a$ and $b$ are coprime. 

\begin{definition}
 A word is \emph{circularly balanced} if all its conjugates (including the word itself) are balanced.
\end{definition}

\begin{proposition}[see for instance~\cite{DBLP:journals/tcs/LucaL06a}]
A word is circularly balanced if and only if it is a conjugate of a Christoffel word, i.e., if and only if it is a row of a Christoffel matrix.
\end{proposition}

So, the conjugates of the lower and upper Christoffel word with Parikh vector $(a,b)$, i.e., the circularly balanced words  with Parikh vector $(a,b)$, are contained in the region $R(a,b)$ of the discrete plane whose contour is delimited by the lower and the upper Christoffel word with Parikh vector $(a,b)$ (Fig.~\ref{fig:stripe}).

 From a geometric point of view, however, \emph{every} balanced word with Parikh vector $(a,b)$ is a digital approximation of the Euclidean segment from $(0,0)$ to $(a,b)$. Indeed, all these words encode paths that are contained in the region $R(a,b)$. A combinatorial proof of this fact is given in the next Proposition \ref{prop:dura}.

It is worth noticing that, in general, there  also exist non-balanced words with Parikh vector $(a,b)$ whose corresponding paths lie in the region $R(a,b)$. As an example, the word $01001001100$ is contained in the region $R(7,4)$ but is not balanced.

\begin{example}
Out of the $\binom{7+4}{4}=330$ binary words with Parikh vector $(7,4)$, only $112$ of them encode paths that lie no more than $\sqrt 2$ away
from the Euclidean segment joining $(0,0)$ to $(7,4)$, i.e., are contained in the region $R(7,4)$ delimited by the lower and the upper Christoffel words of Parikh vector $(7,4)$ 
(see Fig.~\ref{fig:stripe}). In particular, all 19 balanced words with Parikh vector $(7,4)$ 
are among them (see Example~\ref{ex:formula}).
\end{example}


 \begin{proposition}\label{prop:dura}
Let $a,b>0$ and $n=a+b$. For every balanced word $u$ with Parikh vector $(a,b)$, and for every $1<k<n$, the prefix of length $k$ of $u$ has the same Parikh vector as the prefix of length $k$ of either $w_{a,b}$ or $W_{a,b}$.
\end{proposition}

\begin{proof}
By contradiction, let $k$ be the least integer such that there exist a balanced word $x$ with Parikh vector $(a,b)$ whose prefix $u_1$ of length $k$ satisfies either
$|u_1|_1<h$ or $|u_1|_1>h+1$, where $h=|w_{a,b}[1..k]|_1$ (and $h+1=|W_{a,b}[1..k]|_1$). Without loss of generality, assume the former case holds.

Let $n=dk+r$ for some integers $d,r$ with $d\geq 1$ and $0\leq r<k$, so that we may write $x=u_1\cdots u_dv$
for some words such that $|v|=r$ and $|u_i|=k$ for $i=1,\ldots,d$.
Since $x$ is balanced, each of its factors of length $k$ has at most $h$ ones, whereas all factors of the same length in $w_{a,b}$ have at \emph{least} $h$ ones. Together with $|x|_1=q$ and $|u_1|_1<h$, this yields $|u_1\cdots u_d|_1<|w_{a,b}[1..dk]|_1$ and hence $|v|_1>|w_{a,b}[dk+1..n]|_1$. In particular, this implies $r>0$.

Therefore, $v^R$ is a prefix of the balanced word $x^R$ that has more ones than $(w_{a,b}[dk+1..n])^R=(W_{a,b}[1..r])$. As $r<k$, this contradicts our choice for $k$.
\end{proof}

From Proposition~\ref{prop:dura}, we can derive the following characterization of lower Christoffel words, which is somehow dual to that presented in Proposition~\ref{prop:BL}. 

\begin{corollary}\label{cor:surprising}
For every pair $(a,b)$, the lower Christoffel word $w_{a,b}$ is the smallest (in the lexicographic order) balanced word having  Parikh vector $(a,b)$.
\end{corollary}

From Corollary~\ref{cor:surprising} and  Proposition~\ref{prop:BL}, it follows that if $a$ and $b$ are coprime  integers, all Lyndon words with Parikh vector $(a,b)$ are lexicographically smaller than  all balanced words with the same Parikh vector, with the exception of the primitive lower Christoffel word, which is the only word in the intersection of the two sets.



\section{Counting Balanced Words with a Given Parikh Vector}\label{sec:counting}


From the results presented in the previous section, a natural question arises:

\begin{problem}
 Given two positive integers $a$ and $b$, how many balanced words are there with Parikh vector $(a,b)$?
\end{problem}

A recursive formula for  the number $\Bal(a,b)$ of balanced words with Parikh vector $(a,b)$ was given in~\cite{DBLP:journals/dmtcs/BedarideDJR10}. Here, we give an explicit closed formula.


Notice that the number of binary balanced words of \emph{length} $n=a+b$ is known to be 
\begin{equation}\label{eq:St}
1+\sum_{i=1}^{n}(n-i+1)\phi(i),
\end{equation}
where $\phi$ is the Euler totient function, i.e., $\phi(n)$ is the number of positive integers smaller than or equal to $n$ and coprime with $n$ (cf.~\cite{Mig91,Lip82}). But this seems to be of little help for obtaining a bivariate formula depending on specific values of $a$ and $b$.

For simplicity of exposition, we will make use of the infinite periodic word $w^\omega_{a,b}$. Notice that the infinite periodic word $w^\omega_{a,b}$ (resp.~$W^\omega_{a,b}$) is the digital approximation from below (resp.,~from above) of the ray $y=\frac{b}{a}  x$.

\begin{remark}\label{rem:parikh_prefix}
    From the definition of Christoffel word, it  follows that for all lengths $k\geq 0$, the height of the prefix of length $k$ of $w_{a,b}^\omega$  (resp.~of $W_{a,b}^\omega$) is $\left\lfloor k \frac{b} {a+b}\right\rfloor$ (resp.~$\left\lceil k \frac{b} {a+b}\right\rceil$).
\end{remark}

\begin{lemma}
Let $u$ be a balanced word of Parikh vector $(a,b)$, and let $(\alpha,\beta)$ be the Parikh vector of its fractional root $\rho_u$. Then, either:
\begin{itemize}
    \item $\frac{b-1}{a+1}<\frac{\beta}{\alpha} < \frac{b}{a}$, in which case $u$ is a heavy factor of $w_{\alpha,\beta}^\omega$; or
    \item $\frac{b}{a}\le \frac{\beta}{\alpha} < \frac{b+1}{a-1}$, in which case $u$ is a light factor of $w_{\alpha,\beta}^\omega$.
\end{itemize}
\end{lemma}

\begin{proof}
 It is known (see~\cite{DBLP:journals/tcs/LucaL06a}) that $\rho_u$ is a conjugate of the primitive lower Christoffel word $w_{\alpha,\beta}$, so that $u$ is a factor of length $a+b$ of $w_{\alpha,\beta}^\omega$. If $(\alpha,\beta)=(a,b)$, then $u$ is unbordered, i.e., $u=w_{\alpha,\beta}$ or  $u=W_{\alpha,\beta}$, hence $u$ is a light factor by definition. Suppose then $(\alpha,\beta)\neq (a,b)$, i.e., $\beta/\alpha \neq b/a$. If $u$ is a heavy factor (resp.~a light factor) of $w_{\alpha,\beta}^\omega$, then by Proposition~\ref{prop:dura} it has the
    same Parikh vector as the prefix of length $a+b$ of $W_{\alpha,\beta}^\omega$ (resp., of $w_{\alpha,\beta}^\omega$); therefore, by Remark~\ref{rem:parikh_prefix}, we have
    \begin{equation}
    \label{eq:height}
    \left\lceil\frac\beta{\alpha+\beta}(a+b)\right\rceil=b\quad
    \left(\text{resp., }\left\lfloor\frac\beta{\alpha+\beta}(a+b)\right\rfloor=b\right).
    \end{equation}
    Solving for $\beta$ (resp., for $\alpha$), eq.~\eqref{eq:height} is equivalent to
    $\frac{b-1}{a+1}\alpha<\beta < \frac ba\alpha$ (resp., to $\frac{a-1}{b+1}\beta<\alpha < \frac ab\beta$), whence the statement.
\end{proof}

Conversely, all the heavy (resp.~light) factors of length $a+b$ and minimal period $\alpha+\beta$ of the words $w_{\alpha,\beta}^\omega$ such that $\frac{b-1}{a+1} < \frac{\beta}{\alpha} < \frac{b}{a}$ (resp.~$\frac{b}{a} < \frac{\beta}{\alpha} < \frac{b+1}{a-1}$) have Parikh vector $(a,b)$.



The previous result allows us to give the following formula:

\begin{theorem}
\label{thm:formula}
Let $H_{\alpha,\beta}(n)$ (resp.~$L_{\alpha,\beta}(n)$) be the cardinality of the set of heavy (resp.~light) factors of length $n$ of $w^\omega_{\alpha,\beta}$ having minimal period $\alpha+\beta$, for positive $\alpha$ and $\beta$. 
The number of balanced words with Parikh vector $(a,b)$ is given by
\[\Bal(a,b)=\sum_{\alpha=1}^a\;\sum_{\frac{b-1}{a+1}\alpha<\beta < \frac ba\alpha}H_{\alpha,\beta}(a+b)
+\sum_{\beta=1}^b\;\sum_{\frac{a-1}{b+1}\beta<\alpha\leq\frac ab\beta} L_{\alpha,\beta}(a+b)
\]
if $ab\neq 0$, and $\Bal(a,b)=1$ otherwise.
\end{theorem}


In this way, we reduced the problem of computing $\Bal(a,b)$ to that of computing the quantities $H_{\alpha,\beta}(n)$ and $L_{\alpha,\beta}(n)$. We will now show  arithmetic formulas for computing these quantities.

\begin{proposition}
\label{prop:enne}
    Let $N_{\alpha,\beta}(n)$ be the cardinality of the set of factors of length $n$ of $w^\omega_{\alpha,\beta}$ having minimal period $\alpha+\beta$, for positive $\alpha$ and $\beta$. Then
\[N_{\alpha,\beta}(n)=\begin{cases} 0 & \text{ if } \gcd(\alpha, \beta)>1\, \text{ or } n<\alpha+\beta\\ 2(n-\alpha-\beta+1) &\text{ if }n<\alpha+\beta+\min\{\alpha',\beta'\},\\ n-\max\{\alpha',\beta'\}+1 &\text{ if }\alpha+\beta+\min\{\alpha',\beta'\}\leq n<\alpha+\beta+\max\{\alpha',\beta'\},\\ \alpha+\beta &\text{ otherwise,}\end{cases}\]
where $\alpha'$ (resp., $\beta'$) is the multiplicative inverse of $\alpha$ (resp., $\beta$) modulo $\alpha+\beta$.
\end{proposition}
\begin{proof}
The first case is trivial, since all factors of $w_{\alpha,\beta}^\omega$ have the period $(\alpha+\beta)/\gcd(\alpha,\beta)$. If $\alpha$ and $\beta$ are coprime, then by Remark~\ref{rem:periods} the Christoffel word $w_{\alpha,\beta}$ can be factored as $UV$, where $U$ and $V$ are palindromes of length $\alpha'$ and $\beta'$ respectively. Therefore, we need to count factors of length $n$ in $(UV)^\omega$ that contain an occurrence of $UV$ or of $VU=W_{\alpha,\beta}$. It is easy to see that if
$r:=n-\alpha-\beta<\min\{\alpha',\beta'\}=\min\{|U|,|V|\}$, such factors are precisely all factors of length $n=\alpha'+\beta'+r$ contained in the words $(V[1..r])^RUV(U[1..r])$ and $(U[1..r])^RVU(V[1..r])$. This gives $N_{\alpha,\beta}(n)=2(r+1)$ as desired.

If $\alpha'=|U|\leq r<|V|=\beta'$ (resp., $\beta'\leq r<\alpha'$), then a factor of $(UV)^\omega$ of length $n$ contains an occurrence of $UV$ or $VU$ if and only if it occurs as a factor in $(V[1..r])^RUVU(V[1..r])$ (resp., in $(U[1..r])^RVUV(U[1..r])$); thus, $N_{\alpha,\beta}(n)$ equals $|U|+r+1=n-\beta'+1$ (resp., $|V|+r+1=n-\alpha'+1$).

Finally, if $n\geq\alpha+\beta+\max\{\alpha',\beta'\}$, then \emph{all} $\alpha+\beta$ distinct factors of length $n$ in $(UV)^\omega$ contain at least an occurrence of $UV$ or $VU$.
\end{proof}

\begin{example}\label{ex:N}
Let $\alpha=7$, $\beta=3$, so that $\alpha'=3$ and $\beta'=7$, and $n=11$. The $10=\alpha+\beta$ factors of length $11$ of $w_{7,3}^\omega$ are: $00010010010$, $00100010010$, $00100100010$, $00100100100$, 
$01000100100$, $01001000100$, $01001001000$, $10001001001$, 
$10010001001$, $10010010001$. Among them, those of minimal period $10$ are the following four ($4=2(10-\alpha-\beta+1)$):
$00010010010$, $01001001000$, $10001001001$, $10010010001$.

Let $\alpha=5$, $\beta=4$, so that $\alpha'=2$ and $\beta'=7$, and $n=11$. The $9=\alpha+\beta$ factors of length $11$ of $w_{5,4}^\omega$ are: $00101010100$, $01001010101$, $01010010101$, $01010100101$, 
$01010101001$, $10010101010$, $10100101010$, $10101001010$, 
$10101010010$. Among them, those of minimal period $9$ are the following five ($5=10-\max(\alpha'+\beta')+1$):
$00101010100$, $01001010101$, $01010101001$, $10010101010$, 
$10101010010$.
\end{example}

Next, we need to distinguish by height, i.e., between light and heavy factors.
First of all, notice that for every $n\geq 0$, $w_{\alpha,\beta}^\omega$ has exactly $\alpha+\beta$ distinct factors of length $\alpha+\beta+n-1$. Moreover, we have the following

\begin{proposition}
\label{prop:heav}
    Let $\alpha,\beta$ be coprime positive integers,
    and $u$ be any factor of $w_{\alpha,\beta}^\omega$ of length $\alpha+\beta+n-1$ for some $n\geq 0$. The number of occurrences of heavy (resp.~light) factors of length $n$ in $u$ is $n\beta\bmod(\alpha+\beta)$ (resp.~$n\alpha\bmod(\alpha+\beta)$). 
\end{proposition}
\begin{proof}
   We give the proof for heavy factors only, as the one for light factors will follow from it. First notice that $u$ has exactly $\alpha+\beta$ occurrences of factors of length $n$. We prove our claim by induction on $n$. The case $n=0$ is trivial, as the empty word is a light factor by our definition. Let then $n\geq 1$, and assume the statement holds for $n-1$, so that $u=u'x$ for some $x\in\{0,1\}$ and $u'$ has $(n-1)\beta\bmod(\alpha+\beta)$
    occurrences of heavy factors of length $n-1$.

    If the $n$th letter of $w_{\alpha,\beta}^\omega$ is $0$,
    then each occurrence of a heavy factor of length $n-1$ in $u'$ is necessarily followed by $0$, by the balance condition. It follows that the occurrences of heavy factors of length $n-1$ are obtained by extending heavy factors with a $0$ or light factors with a $1$. Since the suffix of $u$ of length $\alpha+\beta$ contains $\beta$ ones, we get
    \begin{equation}
        \label{eq:nbeta}
        (n-1)\beta\bmod(\alpha+\beta)+\beta=n\beta\bmod(\alpha+\beta)
    \end{equation}
    heavy factors of length $n$, where equality in~\eqref{eq:nbeta} is due to fact that the total number of occurrences of factors of length $n$ is $\alpha+\beta$.

    If the $n$th letter of $w_{\alpha,\beta}$ is $1$ instead, then
    each occurrence of a light factor of length $n-1$ in $u'$ must be followed by $1$, so that heavy factors of length $n$ are only obtained by extending heavy factors with a $1$. Since the number  of zeroes in $u[n..\alpha+\beta+n-1]$ is $\alpha$, we obtain $(n-1)\beta\bmod(\alpha+\beta)-\alpha$ occurrences of heavy factors,
    once again yielding our claimed value.
\end{proof}



\begin{proposition}
Let $H_{\alpha,\beta}(n)$ be the cardinality of the set of heavy factors of length $n$ of $w^\omega_{\alpha,\beta}$ having minimal period $\alpha+\beta$, for positive $\alpha$ and $\beta$. Then  
\[
H_{\alpha,\beta}(n)=
\begin{cases}
0 &\text{ if }N_{\alpha,\beta}(n)=0,\\[1.5ex]2\left(\displaystyle\sum_{i=0}^{n-\alpha-\beta}\lceil\sigma(n-i)\rceil+\left\lfloor\sigma i\right\rfloor\right)-\lfloor\sigma n\rfloor N_{\alpha,\beta}(n) &\text{ if }n<\alpha+\beta+\min\{\alpha',\beta'\},\\[3.5ex]
\displaystyle\left(\sum_{i=0}^{n-\alpha'}\lfloor\sigma(n-i)\rfloor+\lceil\sigma i\rceil\right)-\lfloor\sigma n\rfloor N_{\alpha,\beta}(n) &\text{ if }\alpha+\beta+\beta'\leq n<\alpha+\beta+\alpha',\\[3.5ex]
\displaystyle\left(\sum_{i=0}^{n-\beta'}\lceil\sigma(n-i)\rceil+\lfloor\sigma i\rfloor\right)-\lfloor\sigma n\rfloor N_{\alpha,\beta}(n) &\text{ if }\alpha+\beta+\alpha'\leq n<\alpha+\beta+\beta',\\[3.5ex]
n\beta\bmod(\alpha+\beta) &\text{ otherwise,}
\end{cases}
\]
where $\alpha'$ (resp., $\beta'$) is the multiplicative inverse of $\alpha$ (resp., $\beta$) modulo $\alpha+\beta$, and $\sigma=\beta/(\alpha+\beta)$.
\end{proposition}

\begin{proof}
    The first case is clear, as $H_{\alpha,\beta}(n)\leq N_{\alpha,\beta}(n)$ in general. If $\gcd(\alpha,\beta)=1$, then as in the proof of Proposition~\ref{prop:enne} we set $r=n-\alpha-\beta=n-\alpha'-\beta'$ for simplicity, and write $w_{\alpha,\beta}=UV$ for palindromes $U,V$ such that $|U|=\alpha'$ and $|V|=\beta'$.

    If $0<r<\max\{\alpha',\beta'\}$, we can obtain $H_{\alpha,\beta}(n)$ by calculating the sum of all heights of factors of length $n$ and minimal period $\alpha+\beta$, and then subtracting the sum of heights of $N_{\alpha,\beta}(n)$ \emph{light} factors, i.e., $\lfloor\sigma n\rfloor N_{\alpha,\beta}(n)$.

    For the case $r<\min\{\alpha',\beta'\}$, by Proposition~\ref{prop:enne} it suffices to observe that every factor of length $n$ in $(V[1..r])^RUV(U[1..r])$ (and similarly in its reverse
    ) can be split in two parts, the first one ending after the occurrence of $UV$. In other words, for $i=0,\ldots,r$ the two parts are $(VU(V[1..r-i]))^R$ and $U[1..i])$, respectively of heights $\lceil\sigma(n-i)\rceil$ and $\lfloor\sigma i\rfloor$.

    The cases $\beta'\leq r<\alpha'$ and $\alpha'\leq r<\beta'$ are similarly dealt with, by considering the words $(V[1..r])^RUVU(V[1..r])$ and $(U[1..r])^RVUV(U[1..r])$ found in the proof of Proposition~\ref{prop:enne}.

    Finally, in the case where $r\geq\max\{\alpha',\beta'\}$, all $N_{\alpha,\beta}(n)=\alpha+\beta$ factors of length $n$ have minimal period $\alpha+\beta$ and occur exactly once in each factor of $w_{\alpha,\beta}^\omega$ of length $\alpha+\beta+n-1$. The assertion then follows by Proposition~\ref{prop:heav}.
\end{proof}

\begin{example}\label{ex:H}
Let $\alpha=7$, $\beta=3$, so that $\alpha'=3$ and $\beta'=7$, and $n=11$. Among the four factors of length $11$ and minimal period $10=\alpha+\beta$ of  $w_{7,3}^\omega$, namely
$00010010010$, $01001001000$, $10001001001$, $10010010001$ (see Example~\ref{ex:N}), the heavy factors are the last two of the list, while the first two are light.

Let $\alpha=5$, $\beta=4$, so that $\alpha'=2$ and $\beta'=7$, and $n=11$. Among the five factors of length $11$ and minimal period $9=\alpha+\beta$ of  $w_{5,4}^\omega$, namely
$00101010100$, $01001010101$, $01010101001$, $10010101010$, 
$10101010010$ (see Example~\ref{ex:N}), the heavy factors are the last four of the list, while the first one is the only light factor.
\end{example}

Since
\[L_{\alpha,\beta}(n)=N_{\alpha,\beta}(n)-H_{\alpha,\beta}(n),\]
we have all the ingredients to count the number of balanced words with a given Parikh vector.

\begin{example} 
\label{ex:formula}
Let $a=7$, $b=4$. There are 19 balanced words with Parikh vector $(7,4)$, namely: 
$00100100101$, $00100101001$, $00101001001$, $00101001010$, 
$00101010010$, $00101010100$, $01001001001$, $01001001010$, 
$01001010010$, $01001010100$, $01010010010$, $01010010100$, 
$10001001001$, $10010001001$, $10010010001$, $10010010010$, 
$10010010100$, $10010100100$, $10100100100$.
The pairs $(\alpha,\beta)$ with $1\leq \alpha$, $1\leq \beta$ and $\frac{b-1}{a+1}<\frac{\beta}{\alpha} < \frac{b}{a}$ are: $(2,1)$, $(5,2)$, and $(7,3)$; those with $1\leq \alpha$, $1\leq \beta$ and $\frac{b}{a}\le \frac{\beta}{\alpha} < \frac{b+1}{a-1}$ are: $(3,2)$, $(4,3)$, $(5,3)$, $(5,4)$ and $(7,4)$.
Theorem~\ref{thm:formula} gives
\[\begin{split}
\Bal(7,4) &=H_{2,1}(11)+H_{5,2}(11)+H_{7,3}(11)\\&\quad+L_{3,2}(11)+L_{4,3}(11)+L_{5,3}(11)+L_{5,4}(11)+L_{7,4}(11)\\
&=2+1+2+3+2+6+1+2=19\,.
\end{split}\]
\end{example}

\begin{figure}[ht]
    \centering
 \includegraphics[width=300px]{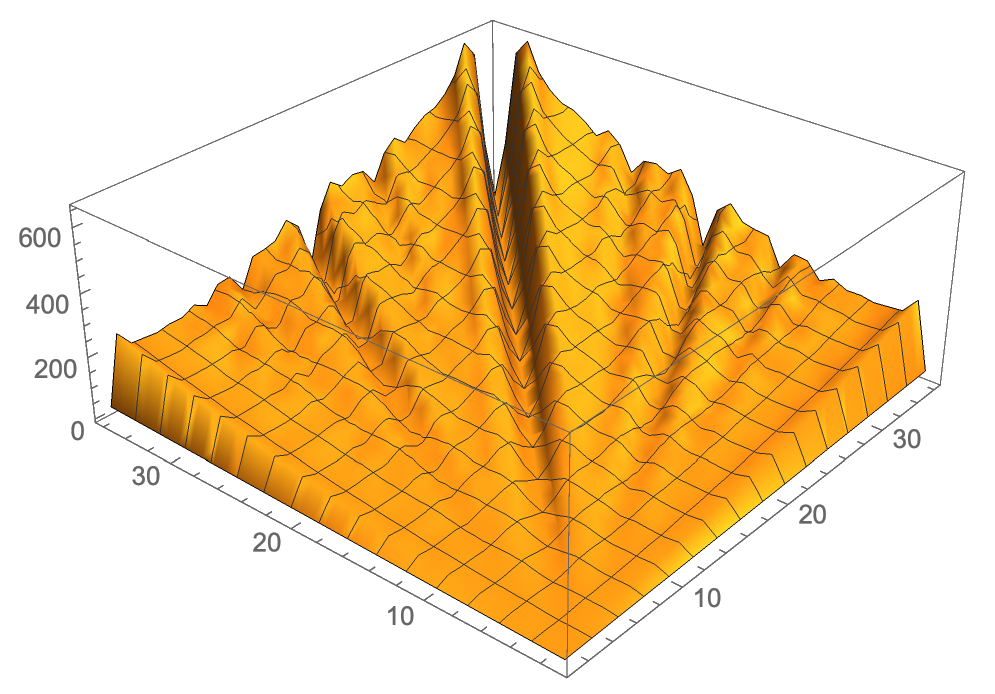}   
    \caption{A three-dimensional plot showing the number of binary balanced words with Parikh vector $(a,b)$ for $a$ and $b$ ranging from $1$ to $36$.}
    \label{fig:plot}
\end{figure}

In Fig.~\ref{fig:plot} we plotted the number of binary balanced words with Parikh vector $(a,b)$ for $a$ and $b$ ranging from $1$ to $36$.

\section{Minimal Forbidden Balanced Words and Minimal Almost Balanced Words}\label{mfw}

In this section, we give a description of those words that are not balanced but such that all of their proper factors are balanced. Geometrically, they represent those paths that have a minimal difference with the paths that are encoded by balanced words.

\begin{definition}
 A balanced word $v$ over $\{0,1\}$ is \emph{right special} (resp.,~\emph{left special}) if both $v0$ and $v1$ are balanced (resp.,~if both $0v$ and $1v$ are balanced). It is \emph{bispecial} if it is both left and right special. A bispecial balanced word $v$ is \emph{strictly bispecial} if all of $0v1$, $1v0$, $0v0$, $1v1$ are balanced~(see~\cite{DBLP:journals/tcs/BerstelL97}).
\end{definition}

Strictly bispecial balanced words are precisely central words~\cite{DBLP:journals/tcs/Luca97a}. Hence, left special (resp.~right special) balanced words are precisely the prefixes (resp.~suffixes) of central words.

\begin{theorem}[\cite{DBLP:journals/jcss/Fici14}]\label{thm:fici}
 A balanced word $v$ is bispecial if and only if $0v1$ is a (not necessarily primitive) lower Christoffel word.
\end{theorem}

In particular, if (and only if) $0v1$ is a \emph{primitive} lower Christoffel word (i.e., $v$ is a palindrome, hence a central word) the word $v$ is a strictly bispecial balanced word.

\begin{example}
 Let $0v1$ be the Christoffel word $0\cdot 0100\cdot 1$. The word $0100$ is bispecial but not strictly bispecial. Indeed, $0\cdot 0100 \cdot 1$, $0\cdot 0100 \cdot 0$ and $1\cdot 0100 \cdot 1$  are balanced, but  $1\cdot 0100 \cdot 0$ is not.
\end{example}

\begin{definition}
Let $L$ be a factorial language, i.e.,  a (finite or infinite) set of words closed under taking factors. We say that a word $w$ is a \emph{minimal forbidden word} for $L$ if $w$ does not belong to $L$ but all proper factors of $w$ do. 
\end{definition}

Let $\MF(L)$ denote the set of minimal forbidden words of the factorial language $L$. A word $w=xvy$, $x,y$ letters, belongs to $\MF(L)$ if and only if
\begin{enumerate}
 \item $xvy\not\in L$;
 \item $xv,vy\in L$.
\end{enumerate}



\begin{theorem}[\cite{DBLP:conf/birthday/MignosiRS99}]
 There is a bijection between factorial languages and their sets of minimal forbidden words.
\end{theorem}

As a consequence, $\MF(L)$ uniquely determines $L$.

The next theorem gives a characterization of the minimal forbidden words for the language $\Bal$ of binary balanced words.

\begin{theorem}[\cite{DBLP:journals/jcss/Fici14}]
 $\MF(\Bal)=\{yvx \mid \{x,y\}=\{0,1\}$, $xvy$ is a non-primitive Christoffel~word$\}$.
\end{theorem}

\begin{example}
The word  $0\cdot 0010\cdot 1$ is not balanced, but all its proper factors are. Indeed, $1\cdot 0010\cdot 0$ is the square of the primitive upper Christoffel word $100$.

The word $0\cdot 0010010\cdot 1$ is not balanced, but all its proper factors are. Indeed, $1\cdot 0010010\cdot 0$ is the cube of the primitive upper Christoffel word $100$.
\end{example}

\begin{corollary}[\cite{DBLP:journals/jcss/Fici14}]\label{cor:Fici}
 For every $n>0$, there are exactly $n-\phi(n)-1$ words of length $n$ in  $\MF(\Bal)$ that start with $0$, and they are all Lyndon words. Here $\phi$ is the Euler totient function.
\end{corollary}


Recall that a word $w$  is not balanced if and only if there exists a palindrome $v$ such that $0v0$ and $1v1$ are both factors of $w$~\cite{LothaireAlg}. We call such a pair of factors an \emph{imbalance pair}.
In 2011, Proven\c{c}al~\cite{DBLP:journals/tcs/Provencal11} studied the language  of \emph{minimal almost balanced words}, $\MAB$s, i.e., minimal (for the factorial order) words with the property that there exists a unique imbalance pair. 

\begin{example}
The word $000101$ is almost balanced with unique imbalance pair $000,101$, but all its proper factors are balanced. Hence it is a $\MAB$.

 On the contrary, the word $000100101$ is not a $\MAB$, since $000,101$ and $000100,100101$ are distinct imbalance pairs. 
\end{example}

\begin{theorem}[\cite{DBLP:journals/tcs/Provencal11}]
$\MAB=\{u^2v^2, (u^2v^2)^R \mid uv$ is the standard factorization of a primitive lower Christoffel word$\}$.
\end{theorem}

We now give an alternative description:

\begin{theorem}
$\MAB
=\{ywx \mid \{x,y\}=\{0,1\}$, $xwy$ is the square of a primitive Christoffel~word$\}$. Therefore, 
$\MAB\subseteq \MF(\Bal)$.
\end{theorem}

\begin{proof}
 Let $uv=0C1$ be the standard factorization of a lower primitive Christoffel word. Let $u=0Q1$, $v=0P1$ for $P,Q$ central words, so that $uv=0Q10P1=0P01Q1$. Then $u^2v^2=0Q10Q10P10P1=0Q10P01Q10P1$, hence $1Q10P01Q10P0=(1C0)^2$. The other cases are similar.
\end{proof}





If $w=xCyxCy$ is a square of a primitive Christoffel word, then the corresponding MAB is $yCyxCx$. A word of this form is in fact a trapezoidal non-balanced word. These words have been studied by D'Alessandro~\cite{DBLP:journals/tcs/DAlessandro02} and by the authors with Michelangelo Bucci~\cite{DBLP:journals/tcs/Bucci0F13}, where words of the form $CyxC$ are called semicentral words.

\end{document}